# Multiplexing Gain of Amplify-Forward Relaying in Wireless Multi-Antenna Relay Networks

Shahab Oveis Gharan and Amir K. Khandani

Coding & Signal Transmission Laboratory
Department of Electrical & Computer Engineering
University of Waterloo
Waterloo, ON, N2L 3G1
shahab, khandani@cst.uwaterloo.ca

**Abstract**

This paper studies the general multi-antenna multiple-relay network. Every two nodes of the network are either connected together through a Rayleigh fading channel or disconnected. We study the ergodic capacity of the network in the high SNR regime. We prove that the traditional amplify-forward relaying achieves the maximum multiplexing gain of the network. Furthermore, we show that the maximum multiplexing gain of the network is equal to the minimum vertex cut-set of the underlying graph of the network, which can be computed in polynomial time in terms of the number of network nodes. Finally, the argument is extended to the multicast and multi-access scenarios.

## I. INTRODUCTION

During the last couple of years, wireless relay networks have received significant attention. This is mainly due to the fact that the relay nodes can potentially enhance the end-to-end coverage and improve the spatial diversity gain. Many different relaying strategies are developed for the relay networks (for example, see [1]–[5]). *Decode-and-Forward* (DF), *Amplify-and-Forward* (AF) and *Compress-and-Forward* (CF) relaying are the main relaying strategies investigated for the wireless relay networks. However, none of the proposed relaying strategies is known to achieve the capacity of the general wireless relay networks. Even in its simplest form which consists of single source, single relay, and single destination, the shannon capacity is unknown.

Among the different relaying strategies, AF relaying turns out to be more suitable in practice. Indeed, in AF relaying the relays are not supposed to decode the transmitted message. Instead, they simply forward their observation of the last time-slot. Hence, the relays consume less computing power. Moreover, the end-to-end system expends a much smaller amount of delay compared with the other relaying strategies, as the relays do not need to wait a couple of time-slots in order to decode the source message or compress the received vector. Another advantage of the AF relaying is that the relay nodes do not need to have any knowledge of the codebook the source is using.

The AF relaying is mainly investigated in literature in order to exploit the cooperative diversity for the wireless relay networks (for example, see [6]–[13]). Indeed, [10] shows that the AF relaying achieves the maximum diversity for any multi-antenna multiple-relay network. Moreover, AF relaying is shown to achieve the optimum diversity-multiplexing trade-off (DMT) in many certain SISO relay networks [8], [10], [13] and also in a number of specific MIMO relay networks [12]. Besides, AF relaying is shown to achieve the capacity of the wireless networks in many asymptotic scenarios [3], [14]–[16]. However, the achievable rate of the AF relaying is unknown for general wireless relay networks. Indeed, [17] has shown that there exists

Financial supports provided by Nortel, and the corresponding matching funds by the Federal government: Natural Sciences and Engineering Research Council of Canada (NSERC) and Province of Ontario: Ontario Centres of Excellence (OCE) are gratefully acknowledged.

scenarios for which the gap between the achievable rate of AF relaying and the capacity of the Gaussian relay network can be as large as possible.

Most recently, Avestimehr *et al.* in [17] show that a variant of the CF relaying achieves the capacity of any general single-antenna gaussian relay network within a constant bit number that only depends on the number of nodes in the network. Furthermore, the authors show in [18] that the result is still valid for both the multi-antenna gaussian and the multi-antenna ergodic Rayleigh fading relay networks. For the case of the relay network with nodes which are equipped with multi-antenna, the gap is only related to the summation of the number of antennas of all network nodes. Also, by relating the original problem to the linear deterministic network and applying the result of [19], the authors of [20] show that the maximum multiplexing gain of the wireless relay networks is equal to the minimum rank between the matrices of different cut-sets of the underlying graph of the network. However, the scheme of [20] also has the drawbacks of CF relaying: each relay node listens for $T$ time-slots ($T$ should approach infinity such that the argument is valid) and then multiplies the received vector by a predefined matrix of size $NT \times NT$, where $N$ is the maximum number of antennas among all nodes of the network and sends the result in the following $T$ time-slots. Hence, the scheme requires high computing power consumption at the relay nodes and imposes a large delay to the end-to-end network.

In this paper, we investigate the potential benefits of traditional AF relaying in the wireless multiple-antenna multiple-relay networks with Rayleigh fading channels. In traditional AF relaying, each relay node forwards its received signal of the last time-slot in the following time-slot. No channel state knowledge is required at either the source or any of the relay nodes. However, the destination is assumed to know the end-to-end channel state. We study the pre-log coefficient of the ergodoc capacity in high SNR regime, known as the multiplexing gain. We prove that the traditional AF relaying achieves the maximum multiplexing gain for any wireless multi-antenna relay network. Furthermore, we characterize the maximum multilexing gain of the network in terms of the minimum vertex cut-set of the underlying graph of the network and show that it can be computed in polynomial-time (with respect to the number of network nodes) using the maximum-flow algorithm. Finally, we show that the argument can be easily extended to the *multicast* and *multi-access* scenarios as well.

The rest of the paper is organized as follows. Section II describes the system model and the main result of the paper. Section III is dedicated to the proof of the main result. Section IV concludes the paper.

*A. Notations*

Throughout the paper, the superscripts $^T$ and $^H$ stand for matrix operations of transposition and conjugate transposition, respectively. Capital bold letters represent matrices, while lowercase bold letters and regular letters represent vectors and scalars, respectively. $\|\mathbf{v}\|$ denotes the norm of vector $\mathbf{v}$ while $\|\mathbf{A}\|$ represents the Frobenius norm of matrix $\mathbf{A}$. $|\mathbf{A}|$ denotes the determinant of matrix $\mathbf{A}$. $\log(.)$ denotes the base-2 logarithm. Motivated by the definition in [21], we define the notation $f(P) \doteq g(P)$ as $\lim_{P \to \infty} \frac{f(P)}{\log(P)} = \lim_{P \to \infty} \frac{g(P)}{\log(P)}$. Similarly, $f(P) \dot{\leq} g(P)$ and $f(P) \dot{\geq} g(P)$ are equivalent to $\lim_{P \to \infty} \frac{f(P)}{\log(P)} \leq \lim_{P \to \infty} \frac{g(P)}{\log(P)}$ and $\lim_{P \to \infty} \frac{f(P)}{\log(P)} \geq \lim_{P \to \infty} \frac{g(P)}{\log(P)}$, respectively.

## II. SYSTEM MODEL AND THE MAIN RESULT

The wireless relay network studied here consists of $K$ relays assisting the source and the destination in the full-duplex mode. Each two nodes are assumed to be either i) connected by a quasi-static flat Rayleigh-fading channel, i.e. the channel gains remain constant during a block of transmission and change independently from block to block; or ii) disconnected, i.e. there is no direct link between them. Hence, the directed graph $G = (V, E)$ is used to show the connected pairs in the network.

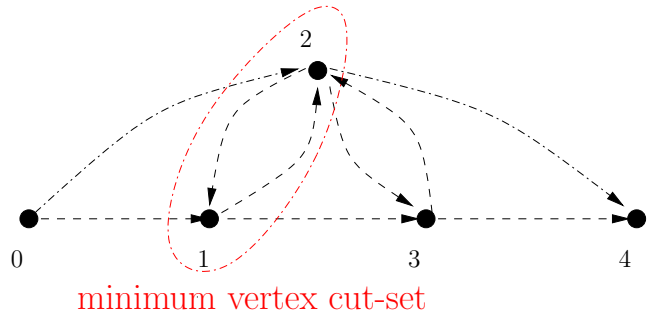

Fig. 1. A wireless multi-antenna relay network. $N_0 = N_4 = 6, N_1 = 3, N_2 = 2, N_3 = 4$. The minimum vertex cut-set is depicted.

The node set is denoted by $V = \{0, 1, \ldots, K+1\}$ where the $i$'th node is equipped with $N_i$ antennas. Nodes $0$ and $K+1$ correspond to the source and the destination nodes, respectively. The received and the transmitted vectors at the $k$'th node are shown by $\mathbf{y}_k$ and $\mathbf{x}_k$, respectively. Hence, at the receiver side of the $a$'th node, we have

$$\mathbf{y}_a = \sum_{(a,b) \in E} \mathbf{H}_{a,b} \mathbf{x}_b + \mathbf{n}_a, \qquad (1)$$

where $\mathbf{H}_{a,b}$ shows the $N_a \times N_b$ Rayleigh-distributed channel matrix between the $a$'th and the $b$'th nodes and $\mathbf{n}_a \sim \mathcal{N}(\mathbf{0}, \mathbf{I}_{N_a})$ is the additive white Gaussian noise. All nodes have the same power constraint, $P$.

In the studied traditional AF relaying, all the relays are always active and, in each time-slot, each relay sends the amplified version of the signal it has received in the last time-slot. In order to state the main argument of the paper, we need the following defenitions.

**Definition 1** *For a network with the connectivity graph $G = (V, E)$, a cut-set on $G$ is defined as a subset $\mathcal{S} \subseteq V$ such that $0 \in \mathcal{S}, K+1 \in \mathcal{S}^c$. The weight of the cut-set corresponding to $\mathcal{S}$, denoted by $w_G(\mathcal{S})$, is defined as*

$$w_G(\mathcal{S}) = \sum_{a \in \mathcal{S}, b \in \mathcal{S}^c, (a,b) \in E} N_a N_b. \qquad (2)$$

**Definition 2** *For a relay network with the directed connectivity graph $G = (V, E)$, a vertex cut-set on $G$ is defined as a subset $\mathcal{C} \subseteq V$ such that any directed path in $G$ from $0$ to $K+1$ intersects with one of the nodes in $\mathcal{C}$. In other words, in the subgraph of $G$ induced[1] by $V - \mathcal{C}$ the destination node $K+1$ is disconnected from the source node, $0$. The capacity of a vertex cut-set is defined as*

$$c_G(\mathcal{C}) = \sum_{v \in \mathcal{C}} N_v. \qquad (3)$$

It should be noted that according to the above definition, the subsets $\{0\}$ and $\{K+1\}$ are vertex cut-sets on $G$.

**Theorem 1** *Consider a general multi-antenna full-duplex relay network with the directed connectivity graph $G = (V, E)$. The traditional AF relaying achieves the maximum multiplexing gain of the network, which is equal to*

$$m_G = \min_{\mathcal{C}} c_G(\mathcal{C}), \qquad (4)$$

*where $\mathcal{C}$ is a vertex cut-set on $G$.*

---

[1]For a graph $G = (V, E)$ and a subset $\mathcal{S} \subseteq V$, the subgraph of $G$ induced by $\mathcal{S}$ is defined as a graph $G_\mathcal{S}$ whose underlying vertex set is $\mathcal{S}$ and any two nodes in $G_\mathcal{S}$ are connected by an edge if and only if the similar nodes in $G$ are connected by an edge.

*Remark 1*- It is worth noting that the maximum multiplexing gain value of every multi-antenna network is computable in polynomial time. Indeed, as it is shown in the proof of Theorem 1, the maximum multiplexing gain of the network is equal to the minimum vertex cut-set of the network graph $G$ or equivalently, the minimum cut of the graph $\hat{G}$ defined in the proof of the Theorem. Noting constructing $\hat{G}$ is feasible in polynomial time, its vertex size is linear with $V$ and also the minimum cut is computable in polynomial time from the Ford-Fulkerson Theorem, we conclude that the maximum multiplexing gain of the network is computable in polynomial time.

Figure 1 shows an example of a wireless multi-antenna relay network. In this network, $N_0 = N_4 = 6, N_1 = 3, N_2 = 2, N_3 = 4$. The vertex cut-set which has the minimum capacity is $\mathcal{C} = \{1, 2\}$ and its associated capacity is equal to $c_G(\mathcal{C}) = 5$. Hence, the maximum multiplexing gain of the network is 5, which is achievable by the traditional AF relaying.

The argument of Theorem 1 can be easily generalized to the *multicast* and *multi-access* scenarios as well. In the multicast scenario, the source aims to send a common message to multiple destinations. In contrast, in the multi-access scenario, multiple source nodes attempt to send their independent messages to the common destination node.

**Theorem 2** *(Multicast Scenario) Consider a general multi-antenna full-duplex relay network with the directed connectivity graph $G = (V, E)$. The source node $s \in V$ aims to send a common message to multiple destinations $t_1, t_2, \ldots, t_M \in V$. The traditional AF relaying achieves the maximum multiplexing gain of the system, which is equal to*

$$m_G^{mc} = \min_{1 \leq i \leq M} m_G(s, t_i), \tag{5}$$

*where $m_G(s, t)$ is the minimum vertex cut-set between $s$ and $t$. In other words, $m_G(s, t) \triangleq \min_{\mathcal{C}} c_G(\mathcal{C})$ over all vertex cut-sets $\mathcal{C}$ between $s$ and $t$.*

*Proof:* The proof is straightforward. First, it should be noted that the ergodic capacity of the multicast problem is less than or equal to the minimum value of the network ergodic capacities between the source and each of the destination nodes. As a result, $m_G^{mc} \leq \min_{1 \leq i \leq M} m_G(s, t_i)$. On the other hand, in the traditional AF relaying investigated in Theorem 1, the relay nodes and the source perform the same operation no matter which node the message is being sent to or what the network connectivity graph is. Hence, the argument of Theorem 1 can be applied for the network between $s$ and each $t_i$. Therefore, the traditional AF relaying achieves the multiplexing gain $m_{AF}^{mc} \geq \min_{1 \leq i \leq sM} m_G(s, t_i)$. This proves the argument of the Theorem. ∎

The following Theorem generalizes the argument of Theorem 1 to the multi-access scenario.

**Theorem 3** *(Multi-Access Scenario) Consider a general multi-antenna full-duplex relay network with the directed connectivity graph $G = (V, E)$. Multiple sender nodes $s_1, s_2, \ldots, s_M \in V$ aim to send independent messages $w_1, w_2, \ldots, w_M$ with the rates $r_1 \log(P), r_2 \log(P), \ldots, r_M \log(P)$ to a common destination node $t \in V$. Let us define the "multiplexing gain region" of the network as the set of all possible $M$-tuples $(r_1, r_2, \ldots, r_M)$ for which the destination can almost surely decode the message of all senders. Then, the traditional AF relaying achieves the optimum multiplexing gain region of the network. Furthermore, the optimum multiplexing gain region of the network is equal to*

$$\mathcal{M}_G^{ma} = \left\{ (r_1, r_2, \ldots, r_M) \, \middle| \, \forall \mathcal{S} \subseteq \{1, 2, \ldots, M\}, \, \sum_{m \in \mathcal{S}} r_m \leq m_G(\mathcal{S}, t) \right\}, \tag{6}$$

*where $m_G(\mathcal{S}, t)$ is the minimum vertex cut-set between $\{s_i \, | i \in \mathcal{S}\}$ and $t$. In other words, $m_G(\mathcal{S}, t) \triangleq \min_{\mathcal{C}} c_G(\mathcal{C})$ over all vertex cut-sets $\mathcal{C}$ between $\{s_i \, | i \in \mathcal{S}\}$ and $t$.*

*Proof:* First, we prove that the optimum multiplexing gain region of the network is a subregion of $\mathcal{M}_G^{ma}$. Next, we prove that the traditional AF relaying achieves all the points that lie in $\mathcal{M}_G^{ma}$. For any subset $\mathcal{S} \subseteq \{1, 2, \ldots, M\}$, we assume that the sender nodes in $\{s_i \,|\, i \in \mathcal{S}\}$ are multiple distributed antennas of a super-node $\hat{s}$ and other sender nodes, i.e. $\{s_i \,|\, i \notin \mathcal{S}\}$, do not interfere on the signals corresponding to $\hat{s}$. Hence, we can apply the argument of Theorem 1 for the multiplexing gain of the network between $\hat{s}$ and $t$. Accordingly, for any $M$-tuples that lies in the optimum multiplexing gain region of the network we have $\sum_{m \in \mathcal{S}} r_m \leq m_G(\mathcal{S}, t)$. Now, we prove that the traditional AF relaying achieves all points that lie in the region $\mathcal{M}_G^{ma}$. Let us consider an arbitrary point $(r_1, r_2, \ldots, r_M) \in \mathcal{M}_G^{ma}$. Let us assume the senders are transmitting independent codewords from independent gaussian codebooks of size $P^{r_1}, P^{r_2}, \ldots, P^{r_M}$, respectively. Each relay node amplifies its received signal of the current time-slot and forwards it in the next time-slot. Let us denote the vectors transmitted by $s_1, s_2, \ldots, s_M$ as $\mathbf{x}_1, \mathbf{x}_2, \ldots, \mathbf{x}_M$, respectively, and the vector received by $t$ as $\mathbf{y}$. Going through the same steps as in the proof of Theorem 1, one can show that the multiplexing gain region of AF relaying is equal to the multiplexing gain region of a multiple-access channel with the equation

$$\mathbf{y} = \sum_{i=1}^{M} \boldsymbol{\mathcal{H}}_i \mathbf{x}_i + \mathbf{n}, \tag{7}$$

where $\boldsymbol{\mathcal{H}}_i$ is a matrix of size $N_t \times N_{s_i}$, corresponding to the end-to-end channel from $s_i$ to $t$, its entries are multivariate polynomials of the channel gains of the network and $\mathbf{n}$ is the white gaussian noise vector of variance 1. The destination performs the jointly typical decoding [22] in order to decide on the transmitted messages. The destination can decode with the error probability approaching 0 iff for any subset $\mathcal{S} \subseteq \{1, 2, \ldots, M\}$, we have

$$\left(\sum_{i \in \mathcal{S}} r_i\right) \log(P) \leq I\left(\mathbf{x}_{\mathcal{S}}; \mathbf{y} \,|\, \mathbf{x}_{\mathcal{S}^c}\right), \tag{8}$$

where $\mathbf{x}_{\mathcal{S}} \triangleq \{\mathbf{x}_i \,|\, i \in \mathcal{S}\}$ and $\mathcal{S}^c \triangleq \{1, 2, \ldots M\} - \mathcal{S}$. Furthermore, from (7), we have

$$I\left(\mathbf{x}_{\mathcal{S}}; \mathbf{y} \,|\, \mathbf{x}_{\mathcal{S}^c}\right) = \mathbb{E}\left\{\log \left|\mathbf{I}_{N_t} + P \sum_{i \in \mathcal{S}} \boldsymbol{\mathcal{H}}_i \boldsymbol{\mathcal{H}}_i^H\right|\right\}. \tag{9}$$

Let us consider the network between the super-node $\hat{s}$ consisting of all nodes $\{s_i | i \in \mathcal{S}\}$ as the sender and $t$ as the destination. Revisiting equations (15) and (34) for the network between $\hat{s}$ and $t$, we conclude[2]

$$\lim_{P \to \infty} \frac{\mathbb{E}\left\{\log \left|\mathbf{I}_{N_t} + P \sum_{i \in \mathcal{S}} \boldsymbol{\mathcal{H}}_i \boldsymbol{\mathcal{H}}_i^H\right|\right\}}{\log P} = m_G(\mathcal{S}, t). \tag{10}$$

Therefore, in the high SNR regime, the constraint in (8) is equivalent to the constraint $\sum_{i \in \mathcal{S}} r_i \leq m_G(\mathcal{S}, t)$. However, this constraint is satisfied as the $M$-tuples $(r_1, r_2, \ldots, r_M)$ lies in the region $\mathcal{M}_G^{ma}$. Hence, the destination can decode the transmitted messages with an error probability vanishing to 0 for any $M$-tuples that lies in $\mathcal{M}_G^{ma}$. This completes the proof. ∎

## III. PROOF OF THEOREM 1

First, we prove the argument for the layered graphs. A graph is called layered if all the paths from the source node to the destination node have the same length. Next, we generalize the argument to any directed graph.

The traditional AF relaying scheme can be described as follows. The source node generates a gaussian codebook with codewords of length $TN_0$ where $N_0$ is the number of antennas at the source. In each time-slot, the source node transmits the corresponding $N_0$ symbols of the codeword. Following that, each relay node observes the power of its received signal in every time-slot. If the power of the received signal of the relay is less than or equal to $P \log(P)$, it amplifies the received signal by $\frac{1}{\sqrt{\log(P)}}$ and transmits the amplified signal in the next time-slot. Denoting the path length from the source to the destination by $l_G$, the destination node $K+1$ receives the transmitted symbol of the source node after $l_G - 1$ time-slots. First, we find a lower-bound on the probability that all the relay nodes are active. Let us consider a relay node $i$. Defining $\mathcal{D}_i$ as the event that the relay node $i$ is active, $\mathbb{P}\{\mathcal{D}_i\}$ can be lower-bounded as

$$\begin{aligned} \mathbb{P}\{\mathcal{D}_i\} &= \mathbb{P}\left\{\mathbb{E}\left\{\|\mathbf{y}_i\|^2\right\} \leq P \log(P)\right\} \\ &\geq \mathbb{P}\left\{P \sum_{(j,i)\in E} \|\mathbf{H}_{j,i}\|^2 + 1 \leq P \log(P)\right\} \end{aligned} \quad (11)$$

---

[2]Here, we used the assumption of layered network in the proof of Theorem 1. However, the argument is yet valid for the general case.

Here, $\mathbf{y}_i$ denotes the received vector of size $N_i$ at the node $i$ and $\mathbf{H}_{j,i}$ denotes the channel from node $j$ to node $i$. Let us define $m_i$ as $m_i \triangleq N_i \sum_{(j,i)\in E} N_j$. Noting that $\sum_{(j,i)\in E} \|\mathbf{H}_{j,i}\|^2$ is a Chi-square random variable with $2m_i$ degree of freedom, we have

$$\begin{aligned} \mathbb{P}\{\mathcal{D}_i\} &\geq 1 - \sum_{k=0}^{m_i-1} \frac{\left(\log(P) - P^{-1}\right)^k}{k!} e^{P^{-1} - \log(P)} \\ &\geq 1 - c_i \frac{(\log(P))^{m_i-1}}{P}, \end{aligned} \quad (12)$$

where $c_i \triangleq e \sum_{k=0}^{m_i-1} \frac{1}{k!}$. In deriving (12), it is assumed $P$ is large enough such that $P \geq 1$. Now, defining $\mathcal{D}$ as the event that all the relay nodes of the network are active, we have

$$\begin{aligned} \mathbb{P}\{\mathcal{D}\} &= \mathbb{P}\left\{\cap_{i=1}^{K} \mathcal{D}_i\right\} \\ &\stackrel{(a)}{\geq} \mathbb{P}\left\{\bigcap_{i=1}^{K} \left\{P \sum_{(j,i)\in E} \|\mathbf{H}_{j,i}\|^2 + 1 \leq P \log(P)\right\}\right\} \\ &\stackrel{(b)}{\geq} 1 - c \frac{\log(P)^d}{P}, \end{aligned} \quad (13)$$

where $c, d \geq 0$ are constants that depend only on the characteristics of the graph $G$. Here, $(a)$ follows from (11) and $(b)$ follows from (12) and the fact that the events $\mathcal{A}_1, \mathcal{A}_2, \ldots, \mathcal{A}_K$ where $\mathcal{A}_i \triangleq \left\{P \sum_{(j,i)\in E} \|\mathbf{H}_{j,i}\|^2 + 1 \leq P \log(P)\right\}$ are independent. From (13), we observe that $\mathbb{P}\{\mathcal{D}\} \sim 1$. Hence, without any loss of generality, we can assume that with probability 1, all the relay nodes are active. In other words, the multiplexing gain of this system is equal to the system in which all the relay nodes are always active and transmit. On the other hand, from the above argument, we know that for all the channels $\mathbf{H}_{j,i}$ with probability 1 we have $\|\mathbf{H}_{j,i}\|^2 \leq \log(P)$. Knowing that for all relay nodes the amplification coefficient is equal to $\frac{1}{\sqrt{\log(P)}}$, we conclude that with probability 1 the power of the equivalent noise at the destination side is less than or equal to a constant that depends only on the topology of the network graph. As a result, the multiplexing gain of the AF relaying is equal to the multiplexing gain of a point-to-point channel whose matrix is equal to the equivalent matrix from the source to the destination. Let us denote the equivalent $N_{K+1} \times N_0$ channel matrix, the source transmitted vector, and the destination received vector by $\mathcal{H}$, $\mathbf{x}$, and $\mathbf{y}$, respectively. Accordingly, the multiplexing gain of the AF relaying is equal to the multiplexing gain of the following channel model

$$\mathbf{y} = \mathcal{H}\mathbf{x} + \mathbf{n} \quad (14)$$

where $\mathbf{n} \sim \mathcal{CN}\left(\mathbf{0}, \mathbf{I}_{N_{K+1}}\right)$. In other words, denoting the multiplexing gain of the AF relaying by $m_{AF}$, we have

$$m_{AF} = \lim_{P \to \infty} \frac{\mathbb{E}\left\{\log\left|\mathbf{I}_{N_0} + P \mathcal{H}^H \mathcal{H}\right|\right\}}{\log(P)}. \quad (15)$$

It should be noted that the entries of $\mathcal{H}$ are multivariate polynomials of the entries of $\{\mathbf{H}_{j,i}\}_{(j,i) \in E}$.

Now, let us construct a graph $\hat{G} = (\hat{V}, \hat{E})$ as follows. Corresponding to each relay node $1 \leq i \leq K$ of the original graph $G$, we add $2N_i$ nodes in $\hat{G}$ and denote them by $a_{i,1}, a_{i,2}, \ldots, a_{i,N_i}$ and $b_{i,1}, b_{i,2}, \ldots, b_{i,N_i}$, respectively. Moreover, for every $1 \leq i \leq K, 1 \leq j \leq N_i$, we add an edge from $a_{i,j}$ to $b_{i,j}$. In other words, $(a_{i,j}, b_{i,j}) \in \hat{E}$. Also, corresponding to the source and destination nodes of $G$, we add $N_0 + N_{K+1} + 2$ nodes to $\hat{G}$ and denote them by $b_{0,1}, b_{0,2}, \ldots, b_{0,N_0}$ and $s$ (corresponding to the source node) and $a_{K+1,1}, a_{K+1,2}, \ldots, a_{K+1,N_{K+1}}$ and $t$ (corresponding to the destination node), respectively. $s$ is connected to $b_{0,j}$'s and also $a_{K+1,j'}$'s are connected to $t$. In other words, $(s, b_{0,j}), (a_{K+1,j'}, t) \in \hat{E}$, for $1 \leq j \leq N_0, 1 \leq j' \leq N_{K+1}$. Finally, corresponding to each pair $(i_1, i_2) \in E$ we have $(b_{i_1,j_1}, a_{i_2,j_2}) \in \hat{E}$ for all possible values of $1 \leq j_1 \leq N_{i_1}$ and $1 \leq j_2 \leq N_{i_2}$.

According to the Ford-Fulkerson Theorem [23], there exists a family of $\nu$ edge-disjoint paths $P \equiv \{p_1, p_2, \ldots, p_\nu\}$ in $\hat{G}$ from $s$ to $t$ where $\nu$ is the min-cut value on $\hat{G}$ from $s$ to $t$. Considering the topology of $\hat{G}$, it is easy to verify that $p_i$'s are also vertex disjoint. To show this fact, it should be noted that for every node $v, v \neq s, t$, we have either $\delta_I(v) \leq 1$ or $\delta_O(v) \leq 1$ where $\delta_I(v)$ and $\delta_O(v)$ denote the incoming and outgoing degree of $v$.

Let us consider the network channels realization in which, for every pair $(i_1, i_2) \in E$, the $(j_1, j_2)$'th entry of the matrix $\mathbf{H}_{i_1,i_2}$ is equal to 1 if one of the paths in P passes through the edge $(b_{i_1,j_1}, a_{i_2,j_2})$. Otherwise, the corresponding entry is equal to 0. More precisely, we have

$$\mathbf{H}_{i_1,i_2}(j_2, j_1) = \begin{cases} 1 & \exists\, 1 \leq v \leq \nu : (b_{i_1,j_1}, a_{i_2,j_2}) \in p_v \\ 0 & \text{oth.w.} \end{cases} \quad (16)$$

For each $1 \leq v \leq \nu$, let us denote the first node after $s$ and the last node before $t$ that the path $p_v$ passes through by $b_{0,\beta_v}$ and $a_{K+1,\gamma_v}$, respectively. Since the paths are vetex disjoint, we have $\beta_v \neq \beta_{v'}$ and $\gamma_v \neq \gamma_{v'}$ for every $v \neq v'$. Moreover, as the paths are vetex disjoint, the equivalent end-to-end channel matrix corresponding to this channel's realization is equal to

$$\mathcal{H}(i_1, i_2) = \begin{cases} 1 & \exists\, v : i_1 = \gamma_v, i_2 = \beta_v \\ 0 & \text{oth.w.} \end{cases} \quad (17)$$

From (17) and knowing that $\gamma_v$'s and $\beta_v$'s are different for different values of $v$ imply that for this realization of network channels, we have

$$\text{Rank}(\mathcal{H}) = \nu. \quad (18)$$

Having (18) and applying Theorem 2.11 of [20], we conclude

$$\lim_{P \to \infty} \frac{\mathbb{E}\left\{\log \left|\mathbf{I}_{N_0} + P\mathcal{H}^H \mathcal{H}\right|\right\}}{\log(P)} \geq \nu. \quad (19)$$

Combining (15) and (19), we have

$$m_{AF} \geq \nu. \quad (20)$$

Now, we prove that $\nu$ is indeed the maximum multiplexing gain of the network. If $\nu = \min(N_0, N_{K+1})$, the argument is valid as the maximum multiplexing gain of the network is less than or equal to the number of antennas at either the source or the destination side. Hence, we only have to prove the argument for the case in which $\nu < \min(N_0, N_{K+1})$.

**Lemma 1** *Consider the graph $\hat{G} = (\hat{V}, \hat{E})$. Assume $\nu < \min(N_0, N_{K+1})$ where $\nu$ is the minimum-cut value over $\hat{G}$ from $s$ to $t$. There exists a cut-set $\mathcal{S} \subseteq \hat{V} - \{t\}$ over $\hat{G}$ of minimum weight ($w_{\hat{G}}(\mathcal{S}) = \nu$) and a vertex cutset $\mathcal{C} \subseteq V - \{0, K+1\}$ of* *minimum capacity over G such that*

$$\hat{E}_{\mathcal{S}} = \bigcup_{v \in \mathcal{C}} \bigcup_{i=1}^{N_v} \{(a_{v,i}, b_{v,i})\}, \qquad (21)$$

where $\hat{E}_{\mathcal{S}}$ denotes the edges that cross the cut-set, i.e. $\hat{E}_{\mathcal{S}} \triangleq \left\{ (u,v) \,\middle|\, (u,v) \in \hat{E}, u \in \mathcal{S}, v \in \mathcal{S}^c \right\}$.

*Proof:* Let us consider a cut-set $\mathcal{S} \subseteq \hat{V} - \{t\}$ over $\hat{G}$ of minimum-value. For every $v \in \hat{V}$, let us define $\Delta_O(v) \triangleq \left\{ (v,u) \,\middle|\, (v,u) \in \hat{E} \right\}$ and $\Delta_I(v) \triangleq \left\{ (u,v) \,\middle|\, (u,v) \in \hat{E} \right\}$. It is easy to verify that we have $|\Delta_O(a_{i,j})| = |\Delta_I(b_{i,j})| = 1$ for all possible values of $i$ and $j$. Furthermore, for a subset $\mathcal{S} \subseteq \hat{V}$, let us define $\mathcal{A}_{\mathcal{S}}$ and $\mathcal{B}_{\mathcal{S}}$ as $\mathcal{A}_{\mathcal{S}} \triangleq \{a_{i,j} \,|\, a_{i,j} \in \mathcal{S}\}$ and $\mathcal{B}_{\mathcal{S}} \triangleq \{b_{i,j} \,|\, b_{i,j} \in \mathcal{S}\}$, respectively. Let us define a new cut-set $\mathcal{T}$ as $\mathcal{T} = \{s\} \cup \mathcal{A}_{\mathcal{T}} \cup \mathcal{B}_{\mathcal{T}}$ where

$$
\begin{aligned}
\mathcal{A}_{\mathcal{T}} &\triangleq \mathcal{A}_{\mathcal{S}} \cup \left\{ v \,\middle|\, v \in \mathcal{A}_{\mathcal{S}^c}, \left|\Delta_I(v) \cap \hat{E}_{\mathcal{S}}\right| \geq 1 \right\}, \\
\mathcal{B}_{\mathcal{T}} &\triangleq \left\{ v \,\middle|\, v \in \mathcal{B}_{\mathcal{S}}, \left|\Delta_O(v) \cap \hat{E}_{\mathcal{A}_{\mathcal{T}} \cup \mathcal{S}}\right| = 0 \right\}.
\end{aligned} \qquad (22)
$$

We prove that $\mathcal{T}$ is also a cut-set of minimum weight. According to the definition of $\mathcal{T}$, we have $\mathcal{A}_{\mathcal{S}} \subseteq \mathcal{A}_{\mathcal{T}}$ and $\mathcal{B}_{\mathcal{T}} \subseteq \mathcal{B}_{\mathcal{S}}$. Now, we have

$$
\begin{aligned}
w_{\hat{G}}(\mathcal{T}) - w_{\hat{G}}(\mathcal{S}) &= \left|\hat{E}_{\mathcal{T}}\right| - \left|\hat{E}_{\mathcal{S}}\right| \\
&= \left(\left|\hat{E}_{\mathcal{A}_{\mathcal{T}} \cup \mathcal{S}}\right| - \left|\hat{E}_{\mathcal{S}}\right|\right) + \left(\left|\hat{E}_{\mathcal{T}}\right| - \left|\hat{E}_{\mathcal{A}_{\mathcal{T}} \cup \mathcal{S}}\right|\right) \\
&\stackrel{(a)}{=} \sum_{v \in \mathcal{A}_{\mathcal{T}} - \mathcal{A}_{\mathcal{S}}} \left(\left|\Delta_O(v) \cap \hat{E}_{\mathcal{A}_{\mathcal{T}} \cup \mathcal{S}}\right| - \left|\Delta_I(v) \cap \hat{E}_{\mathcal{S}}\right|\right) + \\
&\quad \sum_{v \in \mathcal{B}_{\mathcal{S}} - \mathcal{B}_{\mathcal{T}}} \left(\left|\Delta_I(v) \cap \hat{E}_{\mathcal{T}}\right| - \left|\Delta_O(v) \cap \hat{E}_{\mathcal{A}_{\mathcal{T}} \cup \mathcal{S}}\right|\right) \\
&\stackrel{(b)}{\leq} \sum_{v \in \mathcal{A}_{\mathcal{T}} - \mathcal{A}_{\mathcal{S}}} \left(\left|\Delta_O(v) \cap \hat{E}_{\mathcal{A}_{\mathcal{T}} \cup \mathcal{S}}\right| - 1\right) + \sum_{v \in \mathcal{B}_{\mathcal{S}} - \mathcal{B}_{\mathcal{T}}} \left(\left|\Delta_I(v) \cap \hat{E}_{\mathcal{T}}\right| - 1\right) \\
&\stackrel{(c)}{\leq} 0. \qquad (23)
\end{aligned}
$$

Here, $(a)$ follows from the fact that $\mathcal{A}_{\mathcal{S}} \subseteq \mathcal{A}_{\mathcal{T}}$ and $\mathcal{B}_{\mathcal{T}} \subseteq \mathcal{B}_{\mathcal{S}}$ and using the basic arguments of Graph Theory [23] in counting the number of edges of a directed graph. $(b)$ follows from the fact that for every $v \in \mathcal{A}_{\mathcal{T}} - \mathcal{A}_{\mathcal{S}}$ we have $\left|\Delta_I(v) \cap \hat{E}_{\mathcal{S}}\right| \geq 1$, and also for every $v \in \mathcal{B}_{\mathcal{S}} - \mathcal{B}_{\mathcal{T}}$ we have $\left|\Delta_O(v) \cap \hat{E}_{\mathcal{A}_{\mathcal{T}} \cup \mathcal{S}}\right| \geq 1$. Finally, $(c)$ follows from the facts that i) since $\mathcal{A}_{\mathcal{T}} - \mathcal{A}_{\mathcal{S}} \subseteq \mathcal{A}$ for every $v \in \mathcal{A}_{\mathcal{T}} - \mathcal{A}_{\mathcal{S}}$ we have $|\Delta_O(v)| = 1$, and ii) since $\mathcal{B}_{\mathcal{S}} - \mathcal{B}_{\mathcal{T}} \subseteq \mathcal{B}$ for every $v \in \mathcal{B}_{\mathcal{S}} - \mathcal{B}_{\mathcal{T}}$ we have $|\Delta_I(v)| = 1$. (23) proves that $\mathcal{T}$ is also a cut-set of minimum weight over $\hat{G}$.

Now, we prove that there exists a subset $\mathcal{C} \subseteq V - \{0, K+1\}$ such that $\hat{E}_{\mathcal{T}} = \bigcup_{v \in \mathcal{C}} \bigcup_{i=1}^{N_v} \{(a_{v,i}, b_{v,i})\}$. In order to prove, we first show that for every possible value of $j$ we have $b_{0,j} \in \mathcal{T}$ and $a_{K+1,j} \in \mathcal{T}^c$. Since $w_{\hat{G}}(\mathcal{T}) = \nu < N_0$, there exists a value of $j$ such that $b_{0,j} \in \mathcal{T}$. According to the definition of $\mathcal{T}$, we conclude that $\left|\Delta_O(b_{0,j}) \cap \hat{E}_{\mathcal{T}}\right| \stackrel{(a)}{=} \left|\Delta_O(b_{0,j}) \cap \hat{E}_{\mathcal{A}_{\mathcal{T}} \cup \mathcal{S}}\right| = 0$ where $(a)$ follows from the fact that there exists no edge in $\hat{G}$ between the nodes in the subset $\mathcal{B}$, i.e. $(\mathcal{B} \times \mathcal{B}) \cap \hat{E} = \emptyset$. Now, let us assume there exists a value $j'$ such that $b_{0,j'} \in \mathcal{T}^c$. Since $s \in \mathcal{T}$, we have $\left|\Delta_I(b_{0,j'}) \cap \hat{E}_{\mathcal{T}}\right| = 1$. Hence, considering the cutset $\hat{\mathcal{T}} = \mathcal{T} \cup \{b_{0,j'}\}$, we have

$$
\begin{aligned}
w_{\hat{G}}(\hat{\mathcal{T}}) - w_{\hat{G}}(\mathcal{T}) &= \left|\Delta_O(b_{0,j'}) \cap \hat{E}_{\hat{\mathcal{T}}}\right| - \left|\Delta_I(b_{0,j'}) \cap \hat{E}_{\mathcal{T}}\right| \\
&= \left|\Delta_O(b_{0,j}) \cap \hat{E}_{\hat{\mathcal{T}}}\right| - 1 \\
&= \left|\Delta_O(b_{0,j}) \cap \hat{E}_{\mathcal{T}}\right| - 1 \\
&= -1. \qquad (24)
\end{aligned}
$$

(24) contradicts with the assumption that $\mathcal{T}$ is a cut-set of minimum value. Hence, for all possible values of $j$ we have $b_{0,j} \in \mathcal{T}$. Using the same argument, for all possible values of $j$ we have $a_{K+1,j} \in \mathcal{T}^c$. Hence, the edges that cross the cutset $\mathcal{T}$ are either of type $(a_{i,j}, b_{i,j})$, which we call *inner edges*, or of type $(b_{i_1,j_1}, a_{i_2,j_2})$, which we call *outer edges*. Now, we prove that all edges that cross the cutset $\mathcal{T}$ are inner edges. Let us assume an outer edge $(b_{i_1,j_1}, a_{i_2,j_2}) \in \hat{E}_\mathcal{T}$. We have $\left|\Delta_O(b_{i_1,j_1}) \cap \hat{E}_{\mathcal{A}_\mathcal{T} \cup \mathcal{S}}\right| = \left|\Delta_O(b_{i_1,j_1}) \cap \hat{E}_\mathcal{T}\right| \stackrel{(a)}{>} 0$ where $(a)$ follows from the fact that $(b_{i_1,j_1}, a_{i_2,j_2}) \in \hat{E}_\mathcal{T}$. This inequality contradicts with the assumption that $b_{i_1,j_1} \in \mathcal{B}_\mathcal{T}$. Hence, all the edges that cross $\mathcal{T}$ are inner edges.

Finally, we prove the argument of Lemma. Let us define a subset $\mathcal{C} \subseteq V - \{0, K+1\}$ as

$$\mathcal{C} \triangleq \left\{ v \,\Big|\, v \in V - \{0, K+1\}, \exists\, i : (a_{v,i}, b_{v,i}) \in \hat{E}_\mathcal{T} \right\}. \tag{25}$$

We prove that $\hat{E}_\mathcal{T} = \bigcup_{v \in \mathcal{C}} \bigcup_{i=1}^{N_v} \{(a_{v,i}, b_{v,i})\}$,. First, it should be noted that since all the edges that cross the cutset are inner edges, we have $\hat{E}_\mathcal{T} \subseteq \bigcup_{v \in \mathcal{C}} \bigcup_{i=1}^{N_v} \{(a_{v,i}, b_{v,i})\}$. Now, let us assume that $(a_{v,i}, b_{v,i}) \in \hat{E}_\mathcal{T}$ for some $v \in \mathcal{C}$. Accordingly, we have $\left|\Delta_I(b_{v,i}) \cap \hat{E}_\mathcal{T}\right| = 1$. Since $\mathcal{T}$ is a cutset of minimum weight, we conclude that $\left|\Delta_O(b_{v,i}) \cap \hat{E}_{\mathcal{T} \cup \{b_{v,i}\}}\right| > 0$. Hence, for every $1 \leq i' \leq N_v$ we have

$$\left|\Delta_O(b_{v,i'}) \cap \hat{E}_{\mathcal{T} \cup \{b_{v,i'}\}}\right| \stackrel{(a)}{=} \left|\Delta_O(b_{v,i}) \cap \hat{E}_{\mathcal{T} \cup \{b_{v,i}\}}\right| > 0. \tag{26}$$

Here, $(a)$ results from the fact that i) $\Delta_O(b_{v,i}) = \Delta_O(b_{v,i'})$ and ii) there exists no edge between the nodes in $\mathcal{B}$. From (26) and the definition of $\mathcal{T}$, we conclude that for all $1 \leq i' \leq N_v$, we have $b_{v,i'} \in \mathcal{T}^c$. Using the same argument, we conclude that for all $1 \leq i' \leq N_v$ we have $a_{v,i'} \in \mathcal{T}$. As a result, $\bigcup_{i'=1}^{N_v} \{(a_{v,i'}, b_{v,i'})\} \subseteq E_\mathcal{T}$. This proves

$$\hat{E}_\mathcal{T} = \bigcup_{v \in \mathcal{C}} \bigcup_{i=1}^{N_v} \{(a_{v,i}, b_{v,i})\}. \tag{27}$$

Now, we show that $\mathcal{C}$ is a vertex cutset over $G$. Let us assume $\mathcal{C}$ is not a vertex cut-set. Hence, there exists a path $(0, v_1, v_2, \ldots, v_l, K+1)$ where $v_i \in V - \mathcal{C}$ for all possible $v_i$'s. Accordingly, we construct a path P from $s$ to $t$ in $\hat{G}$ as

$$\mathrm{P} \equiv (s, b_{0,1}, a_{v_1,1}, b_{v_1,1}, a_{v_2,1}, b_{v_2,1}, \ldots, a_{v_l,1}, b_{v_l,1}, a_{K+1,1}, t). \tag{28}$$

It is easy to verify that P is a valid path over $\hat{G}$. Furthermore, as for all $v_i$'s we have $v_i \in V - \mathcal{C}$, we conclude $(a_{v_i,1}, b_{v_i,1}) \notin E_\mathcal{T}$. Also, since $(b_{v_i,1}, a_{v_{i+1},1})$ is an outer edge, we have $(b_{v_i,1}, a_{v_{i+1},1}) \notin E_\mathcal{T}$. Hence, P does not cross $\mathcal{T}$. This contradicts with the assumption that $\mathcal{T}$ is a valid cut-set over $\hat{G}$. As a result, $\mathcal{C}$ is a vertex cut-set over $G$.

Finally, we prove that $\mathcal{C}$ is a minimum vertex cut-set over $G$. Let us consider any arbitrary cut-set $\mathcal{C}' \subseteq V - \{0, K+1\}$ over $G$. Let us consider the subgraph of $G$ induced by $V - \mathcal{C}'$ and denote the set of all vertices to whom the source has a directed path by $\mathcal{Q}$. Clearly, since $\mathcal{C}$ is a vertex cut-set, we have $\{K+1\} \notin \mathcal{Q}$. Now, let us define a cut-set $\mathcal{T}'$ over $\hat{G}$ as

$$\mathcal{T}' \triangleq \{0\} \cup \left(\cup_{1 \leq i \leq N_0} \{b_{0,i}\}\right) \cup \left(\cup_{v \in \mathcal{Q}} \cup_{1 \leq i \leq N_v} \{a_{v,i}, b_{v,i}\}\right) \cup \left(\cup_{v \in \mathcal{C}'} \cup_{1 \leq i \leq N_v} \{a_{v,i}\}\right).$$

As there exists no directed path from $s$ to $V - (\mathcal{C}' \cup \mathcal{Q} \cup \{s\})$ in the subgraph of $G$ induced by $V - \mathcal{C}'$, we conclude that there exists no edge from the nodes in $\{0\} \cup \mathcal{Q}$ to the nodes in $V - (\mathcal{C}' \cup \mathcal{Q} \cup \{s\})$. As a result, we have

$$\hat{E}_{\mathcal{T}'} = \bigcup_{v \in \mathcal{C}'} \bigcup_{i=1}^{N_v} \{(a_{v,i}, b_{v,i})\}. \tag{29}$$

Hence, for any vetex cut-set $\mathcal{C}'$ over $G$ we have $\nu \leq c_G(\mathcal{C}')$. Knowing that there exists a vertex cut-set $\mathcal{C}$ such that $\nu = c_G(\mathcal{C})$, we conclude that $\mathcal{C}$ is the minimum vertex cut-set over $G$. This completes the proof of the Lemma. ∎

Applying Lemma 1, for $\nu < \min(N_0, N_{K+1})$ we have $\nu = \min_\mathcal{C} c_G(\mathcal{C})$ where $\mathcal{C}$ is a vertex cut-set on $G$. On the other hand, when $\nu = \min(N_0, N_{K+1})$, we have $\nu \geq \min_\mathcal{C} c_G(\mathcal{C})$. Hence, applying (20) we have

$$m_{AF} \geq \min_\mathcal{C} c_G(\mathcal{C}). \tag{30}$$

Finally, we upper-bound the maximum multiplexing gain of the network. Let us denote the maximum multiplexing gain of the network by $m_G$. Let us consider the vertex cut-set $\mathcal{C}$ with minimum capacity on $G$. In the cases where $\mathcal{C} = \{0\}$ or $\mathcal{C} = \{K+1\}$, we have $m_G \leq \min\{N_0, N_{K+1}\} = c_G(\mathcal{C})$. Let us assume the network is operating during $T$ time-slots. Let us denote the vector that the source transmits from time-slot 1 upto $\tau$ and the vector that the source transmits during the time-slot $\tau$ by $\mathbf{x}^\tau$ and $\mathbf{x}^{(\tau)}$, respectively. Similarly, $\mathbf{y}^\tau$ and $\mathbf{y}^{(\tau)}$ are defined. Furthermore, let us define $\mathbf{x}_\mathcal{C}$ and $\mathbf{y}_\mathcal{C}$ as the vectors that the nodes in $\mathcal{C}$ transmit and receive, respectively. Since $\mathcal{C}$ has the minimum capacity between the vertex cut-sets, the situation where $\mathcal{C} \neq \{0\}$ and $\mathcal{C} \neq \{K+1\}$ implies $\{0, K+1\} \cap \mathcal{C} = \emptyset$. As $\mathcal{C}$ is a vertex cut-set, $(\mathbf{x}, \mathbf{x}_\mathcal{C}, \mathbf{y})$ form a Markov chain. Hence, we have

$$C \stackrel{(a)}{=} \lim_{T \to \infty} \frac{1}{T} \mathbb{E}\left\{I\left(\mathbf{x}^T; \mathbf{y}^T\right)\right\} \stackrel{(b)}{\leq} \lim_{T \to \infty} \frac{1}{T} \mathbb{E}\left\{I\left(\mathbf{x}^T; \mathbf{x}_\mathcal{C}^T\right)\right\} \stackrel{(c)}{\leq} \lim_{T \to \infty} \frac{1}{T} \mathbb{E}\left\{I\left(\mathbf{x}^T; \mathbf{y}_\mathcal{C}^T\right)\right\}, \tag{31}$$

where $C$ is the ergodic capacity of the network and the operator $\mathbb{E}$ is performed over all channels' realizations. Here, $(a)$ follows from the Fano inequality [22], $(b)$ follows from the fact that $(\mathbf{x}, \mathbf{x}_\mathcal{C}, \mathbf{y})$ form a Markov chain, and $(c)$ follows from the fact that $(\mathbf{x}, \mathbf{y}_\mathcal{C}, \mathbf{x}_\mathcal{C})$ form a Markov chain. Now, $\mathbb{E}\left\{I\left(\mathbf{x}^T; \mathbf{y}_\mathcal{C}^T\right)\right\}$ can be upper-bounded as

$$
\begin{aligned}
\mathbb{E}\left\{I\left(\mathbf{x}^T; \mathbf{y}_\mathcal{C}^T\right)\right\} &\stackrel{(a)}{\leq} \sum_{v \in \mathcal{C}} \mathbb{E}\left\{h\left(\mathbf{y}_v^T\right)\right\} - \mathbb{E}\left\{h\left(\mathbf{y}_\mathcal{C}^T \mid \mathbf{x}^T\right)\right\} \\
&= \sum_{v \in \mathcal{C}} \mathbb{E}\left\{h\left(\mathbf{y}_v^T\right)\right\} - \sum_{\tau=1}^T \mathbb{E}\left\{h\left(\mathbf{y}_\mathcal{C}^{(\tau)} \mid \mathbf{y}_\mathcal{C}^{\tau-1}, \mathbf{x}^T\right)\right\} \\
&\stackrel{(b)}{\leq} \sum_{v \in \mathcal{C}} \mathbb{E}\left\{h\left(\mathbf{y}_v^T\right)\right\} - \sum_{\tau=1}^T \mathbb{E}\left\{h\left(\mathbf{n}_\mathcal{C}^{(\tau)}\right)\right\} \\
&\stackrel{(c)}{\leq} \sum_{v \in \mathcal{C}} \sum_{\tau=1}^T \mathbb{E}\left\{h\left(\mathbf{y}_v^{(\tau)}\right) - h\left(\mathbf{n}_v^{(\tau)}\right)\right\} \\
&\stackrel{(d)}{\leq} T\left(\sum_{v \in \mathcal{C}} N_v\right) \log(P) + TO(1) \\
&= T c_G(\mathcal{C}) \log(P) + TO(1). \tag{32}
\end{aligned}
$$

Here, $(a)$ follows from the fact that $h\left(\mathbf{y}_\mathcal{C}^T\right) \leq \sum_{v \in \mathcal{C}} h\left(\mathbf{y}_v^T\right)$, $(b)$ follows from the fact that $\mathbf{n}_\mathcal{C}^{(\tau)}$ is independent from $\left(\mathbf{x}^T, \mathbf{y}_\mathcal{C}^{\tau-1}, \mathbf{y}_\mathcal{C}^{(\tau)} - \mathbf{n}_\mathcal{C}^{(\tau)}\right)$ and applying entropy power inequality[3] [22], and $(c)$ follows from the fact that $h\left(\mathbf{n}_\mathcal{C}^{(\tau)}\right) = \sum_{v \in \mathcal{C}} h\left(\mathbf{n}_v^{(\tau)}\right)$ and $h\left(\mathbf{y}_v^T\right) \leq \sum_{\tau=1}^T h\left(\mathbf{y}_v^{(\tau)}\right)$. In order to prove $(d)$, let us define $v^-$ as the set of vertices from whom there exsits an edge to $v$, i.e. $v^- \triangleq \{u \mid (u,v) \in E\}$. We have $\mathbb{E}\left\{h\left(\mathbf{y}_v^{(\tau)}\right) - h\left(\mathbf{n}_v^{(\tau)}\right)\right\} = \mathbb{E}\left\{I\left(\mathbf{x}_{v^-}^{(\tau)}; \mathbf{y}_v^{(\tau)}\right)\right\}$, which is equal to the ergodic capacity of a $c_G(v^-) \times N_v$ MIMO system. As a result $\mathbb{E}\left\{h\left(\mathbf{y}_v^{(\tau)}\right) - h\left(\mathbf{n}_v^{(\tau)}\right)\right\} = \min\left(c_G(v^-), N_v\right) \log(P) + O(1) \leq N_v \log(P) + O(1)$, which results in $(d)$. Combining (31) and (32), we have

$$m_G \leq \min_\mathcal{C} c_G(\mathcal{C}). \tag{33}$$

---

[3]According to the entropy power inequality, for any independent random vectors $\mathbf{a}$ and $\mathbf{b}$ of size $n$ we have $2^{\frac{2}{n}h(\mathbf{a}+\mathbf{b})} \geq 2^{\frac{2}{n}h(\mathbf{a})} + 2^{\frac{2}{n}h(\mathbf{b})}$. As a result $h(\mathbf{a}+\mathbf{b}) \geq h(\mathbf{a})$.

Comparing (30) and (33), we conclude

$$m_G = m_{AF} = \min_{\mathcal{C}} c_G(\mathcal{C}). \tag{34}$$

(34) completes the proof of the Theorem for the case of the layered networks.

Now, we prove the argument of the Theorem for the case of any arbitrary networks. First, it should be noted that the inequality series (31) and (32) are still valid for any arbitrary network. As a result, (33) is still valid. Hence, we just need to prove $m_{AF} \geq \min_{\mathcal{C}} c_G(\mathcal{C})$.

In the traditional AF relaying, the network is operated through time-slots $t = 1, 2, \ldots, T$ as follows. The source sends a codeword of length $T$ from its gaussian codebook. Each relay node amplifies its received signal from the last time-slot and forwards it in the next time-slot with the possible amplification coefficient $\frac{1}{\sqrt{\log(P)}}$, similar to what explained for the layered network. The destination decodes the transmitted message using the joint decoding of its received vector from all of its antennas during the time-slots $t = 1, 2, \ldots, T$. Noting the destination has $N_{K+1}$ antennas, its received vector is of size $TN_{K+1}$. Let us denote the transmitted vector at the source and the received vector at the destination by $\mathbf{x} = \mathbf{x}_{t,n_1}$ and $\mathbf{y} = \mathbf{y}_{t,n_2}$, respectively, where $1 \leq t \leq T$, $1 \leq n_1 \leq N_0$ and $1 \leq n_2 \leq N_{K+1}$. Using the same argument we applied for the layered network, we conclude that the multiplexing gain of the AF relaying is equal to the multiplexing gain of a point-to-point MIMO channel whose matrix is of size $TN_{K+1} \times TN_0$ and its entries are multivariate polynomials of the entries of the network channels matrices $\{\mathbf{H}_{j,i}\}_{(j,i) \in E}$. Let us denote this channel matrix by $\boldsymbol{\mathcal{H}} = \boldsymbol{\mathcal{H}}((t_2, n_2), (t_1, n_1))$ where $1 \leq t_1, t_2 \leq T$, $1 \leq n_1 \leq N_0$, and $1 \leq n_2 \leq N_{K+1}$. In other words, we have

$$m_{AF} = \frac{1}{T} \lim_{P \to \infty} \frac{\mathbb{E}\left\{\log\left|\mathbf{I}_{N_0} + P\boldsymbol{\mathcal{H}}^H\boldsymbol{\mathcal{H}}\right|\right\}}{\log(P)}. \tag{35}$$

Here, the expectation is performed over all network channels realization. Now, let us consider the corresponding graph $\hat{G} = (\hat{V}, \hat{E})$, which was previously defined for the unlayered network. It can be shown that the entries of $\boldsymbol{\mathcal{H}}$ are related to the weight of paths in $\hat{G}$ as follows.

$$\boldsymbol{\mathcal{H}}((t_2, n_2), (t_1, n_1)) = \sum_{\mathrm{p}} w(\mathrm{p}),$$
$$\text{s.t. } \mathrm{p}(1) = b_{0,n_1} \ \& \ \mathrm{p}(l(\mathrm{p}) - 1) = a_{K+1,n_2} \ \& \ l(\mathrm{p}) = 3 + 2(t_2 - t_1). \tag{36}$$

Here, the summation is over the weight of all paths p of length[4] $3 + 2(t_2 - t_1)$ in $\hat{G}$ from $s$ to $t$ such that $\mathrm{p}(1) = b_{0,n_1}$ and $\mathrm{p}(l(\mathrm{p}) - 1) = a_{K+1,n_2}$. Furthermore, the weight of a path p is defined as

$$w(\mathrm{p}) \triangleq \prod \mathbf{H}_{i_1,i_2}(j_2, j_1),$$
$$\text{s.t. } (b_{i_1,j_1}, a_{i_2,j_2}) \in \hat{E} \ \& \ \mathrm{p} \text{ passes through } (b_{i_1,j_1}, a_{i_2,j_2}). \tag{37}$$

Applying the same argument as for the layered network, there exists a family of $\nu$ vertex-disjoint paths $\mathrm{P} \equiv \{\mathrm{p}_1, \mathrm{p}_2, \ldots, \mathrm{p}_\nu\}$ in $\hat{G}$ from $s$ to $t$ where $\nu$ is the min-cut value on $\hat{G}$ from $s$ to $t$. Now, let us consider the network channels realization in which for every pair $(i_1, i_2) \in E$, the $(j_2, j_1)$'th entry of the matrix $\mathbf{H}_{i_1,i_2}$ is equal to 1 if one of the paths in P passes through the edge $(b_{i_1,j_1}, a_{i_2,j_2})$, and otherwise the corresponding entry is equal to 0. More precisely, we have

$$\mathbf{H}_{i_1,i_2}(j_2, j_1) = \begin{cases} 1 & \exists \, v : \mathrm{p}_v \text{ passes through } (b_{i_1,j_1}, a_{i_2,j_2}) \\ 0 & \text{oth.w.} \end{cases} \tag{38}$$

---

[4] The length of a path p, which is denoted by $l(\mathrm{p})$, is defined as the number of edges that the path goes through.

From (37) and (38) and knowing the fact that the paths are vertex disjoint, we conclude that for every path p in $\hat{G}$ from $s$ to $t$, we have

$$w(\text{p}) = \begin{cases} 1 & \exists\, 1 \leq v \leq \nu : \text{p} = \text{p}_v \\ 0 & \text{oth.w.} \end{cases} \tag{39}$$

For each $1 \leq v \leq \nu$, let us denote the first node after $s$ and the last node before $t$ that the path $\text{p}_v$ passes through by $b_{0,\beta_v}$ and $a_{K+1,\gamma_v}$, respectively. Since the paths are vetex disjoint, we have $\beta_v \neq \beta_{v'}$ and $\gamma_v \neq \gamma_{v'}$ for every $v \neq v'$. Applying this fact, (36), and (39), we conclude that the equivalent end-to-end channel matrix corresponding to this specific realization for the network channels is equal to

$$\mathcal{H}((t_2, n_2), (t_1, n_1)) = \begin{cases} 1 & \exists\, v : n_1 = \beta_v, n_2 = \gamma_v, l(\text{p}_v) = 2(t_2 - t_1) + 3 \\ 0 & \text{oth.w.} \end{cases}. \tag{40}$$

From (40) and knowing that $\gamma_v \neq \gamma_{v'}$ and $\beta_v \neq \beta_{v'}$ for every $v \neq v'$, we have

$$\text{Rank}(\mathcal{H}) = \sum_{v=1}^{\nu}\left(T - \frac{l(\text{p}_v) - 3}{2}\right) \geq \nu\left(T - \frac{l_{\hat{G}} - 3}{2}\right) = \nu\left(T - l_G + 1\right), \tag{41}$$

where $l_{\hat{G}}$ and $l_G$ denote the maximum length of a simple path[5] connecting the source to the destination in $G$ and $\hat{G}$, respectively. Having (41) and applying Theorem 2.11 of [20], we conclude

$$\lim_{P \to \infty} \frac{\mathbb{E}\left\{\log\left|\mathbf{I}_{N_0} + P\mathcal{H}^H\mathcal{H}\right|\right\}}{\log(P)} \geq \nu\left(T - l_G + 1\right). \tag{42}$$

Combining (35) and (42), we have

$$m_{AF} \geq \nu - \frac{\nu(l_G - 1)}{T}. \tag{43}$$

Applying Lemma 1, for $\nu < \min(N_0, N_{K+1})$, we have $\nu = \min_{\mathcal{C}} c_G(\mathcal{C})$ where $\mathcal{C}$ is a vertex cut-set on $G$. On the other hand, when $\nu = \min(N_0, N_{K+1})$, we have $\nu \geq \min_{\mathcal{C}} c_G(\mathcal{C})$. Hence, applying (43), we have

$$m_{AF} \geq \min_{\mathcal{C}} c_G(\mathcal{C}) - \frac{v(l_G - 1)}{T}, \tag{44}$$

where $\mathcal{C}$ is a vertex cut-set on $G$. Having $T \to \infty$ completes the proof of the Theorem.

## IV. Conclusion

A general wireless multi-antenna multiple-relay network is investigated. Every two nodes of the network are either connected together through a Rayleigh fading channel or disconnected. The ergodic capacity of the network is studied in the high SNR regime. It is shown that the traditional AF relaying achieves the maximum multiplexing gain of the network. Furthermore, the maximum multiplexing gain of the network is proved to be equal to the minimum vertex cut-set of the underlying graph of the network, which can be computed in polynomial time in terms of the number of network nodes. Finally, the argument is extended to the muticast and multi-access scenarios.

---

[5]A path p in a graph $G = (V, E)$ is called simple, if it passes through each vertex of $V$ just once.